\documentclass[aps,prl,showpacs,twocolumn,floatfix,longbibliography]{revtex4-1}
\usepackage{graphicx}
\usepackage{amssymb}
\usepackage{subfigure}
\usepackage{array}
\usepackage{tabularray}

\usepackage{color} 
\usepackage{xcolor}

\newcommand{\ket}[1]{\vert #1 \rangle}

\begin{document}

\title{A millimeter-wave atomic receiver}
\author{R. Legaie}
\author{G. Raithel}
\author{David A. Anderson}
\email{Dave@RydbergTech.com}
\affiliation{Rydberg Technologies Inc., Ann Arbor, MI 48103}


\begin{abstract}

Rydberg quantum sensors are sensitive to radio-frequency fields across an ultra-wide frequency range spanning megahertz to terahertz electromagnetic waves resonant with Rydberg atom dipole transitions. Here we demonstrate an atomic millimeter-wave heterodyne receiver employing continuous-wave lasers stabilized to an optical frequency comb. We characterize the atomic receiver in the W-band at signal frequency of $f$~=~95.992512~GHz, and demonstrate a sensitivity of 7.9~$\mu$V/m/$\sqrt{\mathrm{Hz}}$ and a linear dynamic range of 70~dB. We develop frequency selectivity metrics for atomic receivers and demonstrate their use in our millimeter-wave receiver, including signal rejection levels at signal frequency offsets $\Delta f/f$ = 10$^{-4}$, 10$^{-5}$ and 10$^{-6}$, 3-dB, 6-dB, 9-dB and 12-dB bandwidths, filter roll-off, and shape factor analysis. Our work represents an important advance towards future studies and applications of atomic receiver science and technology and in weak millimeter-wave and high-frequency signal detection.

\end{abstract}

\maketitle

\section{Introduction}

Over the last decade, Rydberg-atom based sensors have shown to be a prolific and competitive platform for radio-frequency~(RF) sensing and for detecting RF electric ($E$)-fields resonant with Rydberg-atom electric-dipole transitions over a broad frequency range from DC to THz~\cite{Sedlacek.2012, Anderson.2014, anderson2.2017, Anderson2.2018, Anderson.2019, Fan.2015, Kumar.2016, Kumar.2017, Miller.2016, Anderson.2016, Anderson.2017, Meyer.2020, Downes.2020, Kunz.2023}, with self-calibration and SI-traceability capability~\cite{Holloway2.2014, HollowayEMC.2017, HollowayAPL.2018}. Atomic field sensor devices have been developed~\cite{RFMSIEEE.2019} and implemented for applications including near-field and wide-area antenna measurement and imaging~\cite{Cardman.2020} and atom radio reception~\cite{AndersonRadioIEEE.2021}.  The combination of electromagnetically induced transparency~(EIT)~\cite{Harris.1990, Fleischhauer.2005, Mohapatra.2007} and spectroscopic readout of field-induced Rydberg level changes such as Autler-Townes~(AT) splittings and AC shifts for atomic electric field determination and measurement~\cite{anderson2.2017} allows Rydberg-atom electrometry to outperform traditional methods of measuring RF~field properties, such as amplitude~\cite{Jing.2020, Xiao.2022}, frequency~\cite{GordonPhase.2019}, phase~\cite{SimonsPhase.2019, Jing.2020, Anderson.2022}, polarization~\cite{Sedlacek.2013, Anderson2.2018} and angle of arrival~\cite{Robinson.2021}.  The measurement of small $E$-field amplitudes with ultra-high sensitivity is critical in metrology and sensing applications, as well as in understanding bandwidth and selectivity performance characteristics of atomic receivers. Atomic receiver selectivity, despite being essential to establish the performance of atomic sensors, has heretofore not been explored. Further, although sensitivity measurements of microwave (MW) $E$-fields around 10~GHz carrier frequencies have been progressing~\cite{Jing.2020, Prajapati.2021, Cai.2022}, and ~V/m field measurements have been conducted at 100~GHz with vapors in subwavelength-imaging~\cite{Gordon.2014, Holloway.2014} and equipment testing~\cite{Thaicharoen.2019}, atomic heterodyne receivers at higher frequencies in the millimeter-wave (MMW) band reaching 100~GHz and above have not been reported, nor have receiver performance metrics including sensitivity, selectivity, bandwidth, and dynamic range. The selectivity of a traditional radio receiver is a measure of the receiver's ability to reject unwanted signals that are at frequencies near the channel in use.  In radio applications receiver selectivity dictates achievable channel bandwidths and the quality of signals received based on rejection levels of signals on adjacent channels. Achieving high receiver selectivity is particularly critical when receiver operation requires mitigation of blanketing, {\sl{i.e.}} receiver blocking by strong unwanted signals, and other intentional or unintentional electromagnetic interference. Applications of millimeter-wave atomic receivers requiring highly sensitive and selective sensors include high-frequency ($>$~100~GHz) wideband wireless communications, next-generation mobile devices~(6G), millimeter-wave imaging devices or aeronautics.

In this work we demonstrate a MMW atomic receiver. The atomic receiver incorporates a cesium vapor detector using two-photon Rydberg electromagnetically induced transparency (EIT) with lasers locked to an optical frequency comb~(OFC) for reduced optical noise and optimized operating laser parameters. The receiver implements a MMW heterodyne (HET) architecture with a local oscillator (LO) reference field providing phase- and frequency-sensitivity and increased atomic receiver responsivity to signal fields.
A minimum field sensitivity of 7.9$~\mu$V/m/$\mathrm{\sqrt{Hz}}$ and a linear dynamic range of $>$~70~dB are demonstrated for a signal field frequency $f_{\mathrm{{SIG}}}~=95.992512$~GHz,
which is resonant with the $\mathrm{37^{2}S_{1/2}}$ to $\mathrm{36^{2}P_{3/2}}$~MMW~transition of Cs.
We investigate the selectivity of atomic apertures for MMW sensing.  Atomic aperture receiver selectivity arises from signal filtering in the analog front-end afforded by the physics principles of the atom-field interaction that occurs in the field sensing element. The filtering in the atom-optical domain occurs prior to any electronic filtering analog or digital signal processing. The receiver selectivity to the signal field is characterized by measurements of the atomic response to MMW signals offset by up to $\Delta f = \pm 15$~MHz relative to the LO.

\begin{figure*}
\includegraphics[width=17cm]{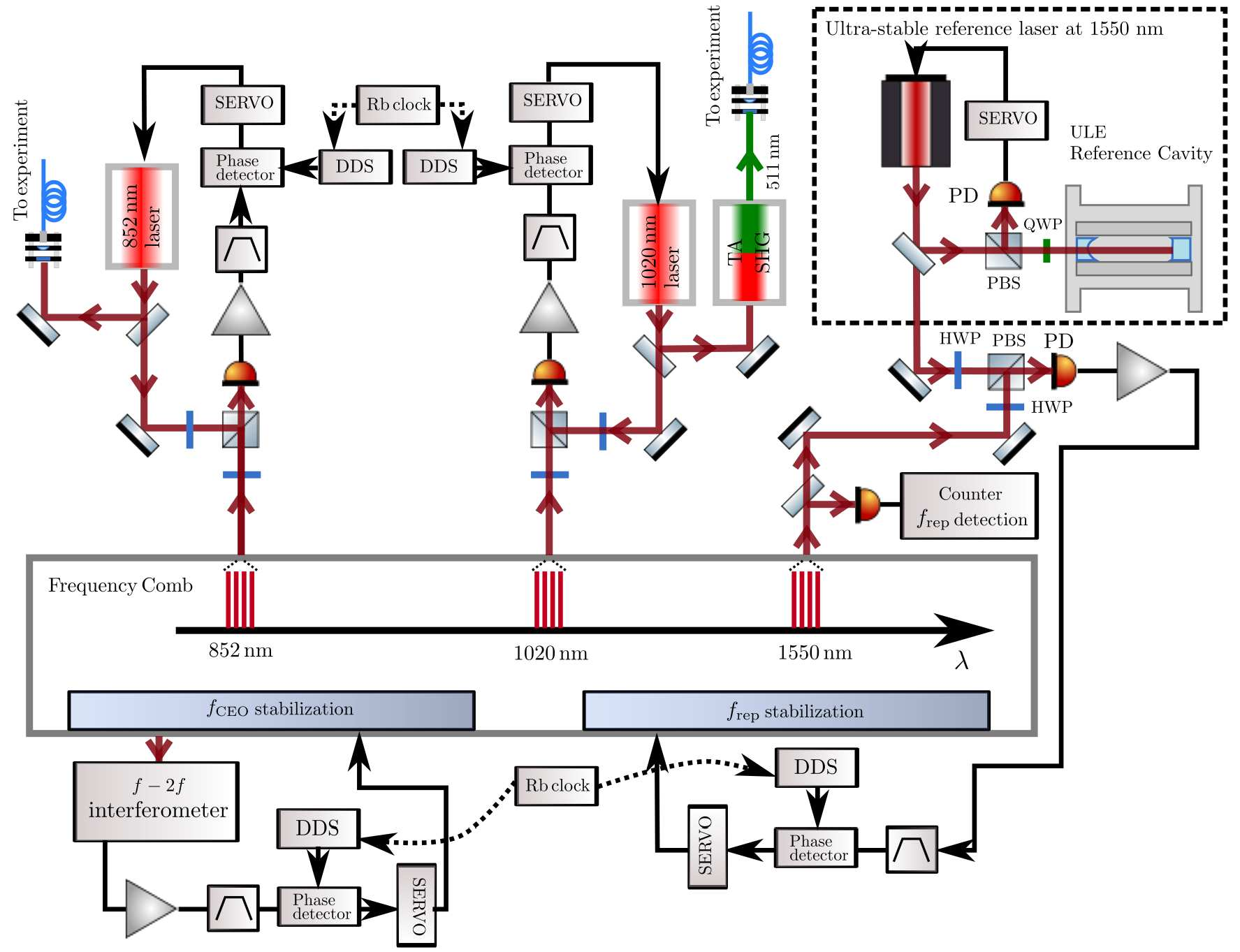}
\caption{Diagram of the optical frequency comb (OFC) stabilization to an ultra-stable reference laser as well as 852-nm and 1020-nm lasers lock to the OFC. HWP: half-waveplate, QWP: quarter-waveplate, PD: photodiode, PBS: polarizing beam-splitter, DDS: direct digital synthesizer, Rb clock:  Rubidium clock. }
\label{fig:1}
\end{figure*}

\section{Rydberg atom sensor with optical frequency comb laser stabilization}
\label{OFC}

A primary design requirement for atomic sensors to perform their intended application functions, including reaching state-of-the-art sensitivity levels, is to account for and address noise sources that may limit the detection of RF $E$-fields with the atoms.
In atomic radio receivers, the noise sources fall into two main categories, namely noise occurring in the front-end atomic vapor cell element, and noise from system hardware and electronics. The latter category includes noise from optical subsystems, excess noise from the photodiode detectors and amplifiers used to read out the EIT signal, and excess noise in the downstream electronic signal processing. The former category includes interactions of the sensor atoms in the vapor cell with incoherent electromagnetic fields, such as blackbody thermal radiation as well as radio-frequency noise, with fields present within the vapor cell, and with other atoms and background gases.
Additional noise may arise from unwanted coherent electromagnetic fields entering the cell, as well as from inhomogeneities of electric fields to be measured
and of tuning fields.

An orthogonal categorization of noise relates to whether a noise source is technical or fundamental in nature. The net noise level of the entire system is often set by the sum of all technical noise sources. However, there are fundamental (physical) limits that include atomic shot noise limits, optical shot noise limits for the weak EIT probe light, single-atom EIT dynamics, atom-field interaction times and Doppler shifts in the atomic vapor cell, black-body fields, and electronic shot noise and Johnson noise limits.

On the laser side, the fundamental noise set by the Schawlow-Townes limit~\cite{Schawlow.1958}
is outweighed by technical, non-fundamental noise. A critical technical limitation of hardware in atom-based quantum devices is the excess frequency jitter and amplitude noise from the lasers sources used to interrogate the quantum states of the atoms. In Rydberg sensing based on atomic vapors where multi-wavelength (e.g. two-, three-, and four-photon) Rydberg EIT approaches are typically implemented, the technical frequency and amplitude noise of each laser source can limit the achievable atomic sensor sensitivity to electromagnetic fields as well as the fidelity of radio receivers derived from such sensors.

\begin{figure*}
\includegraphics[width=17cm]{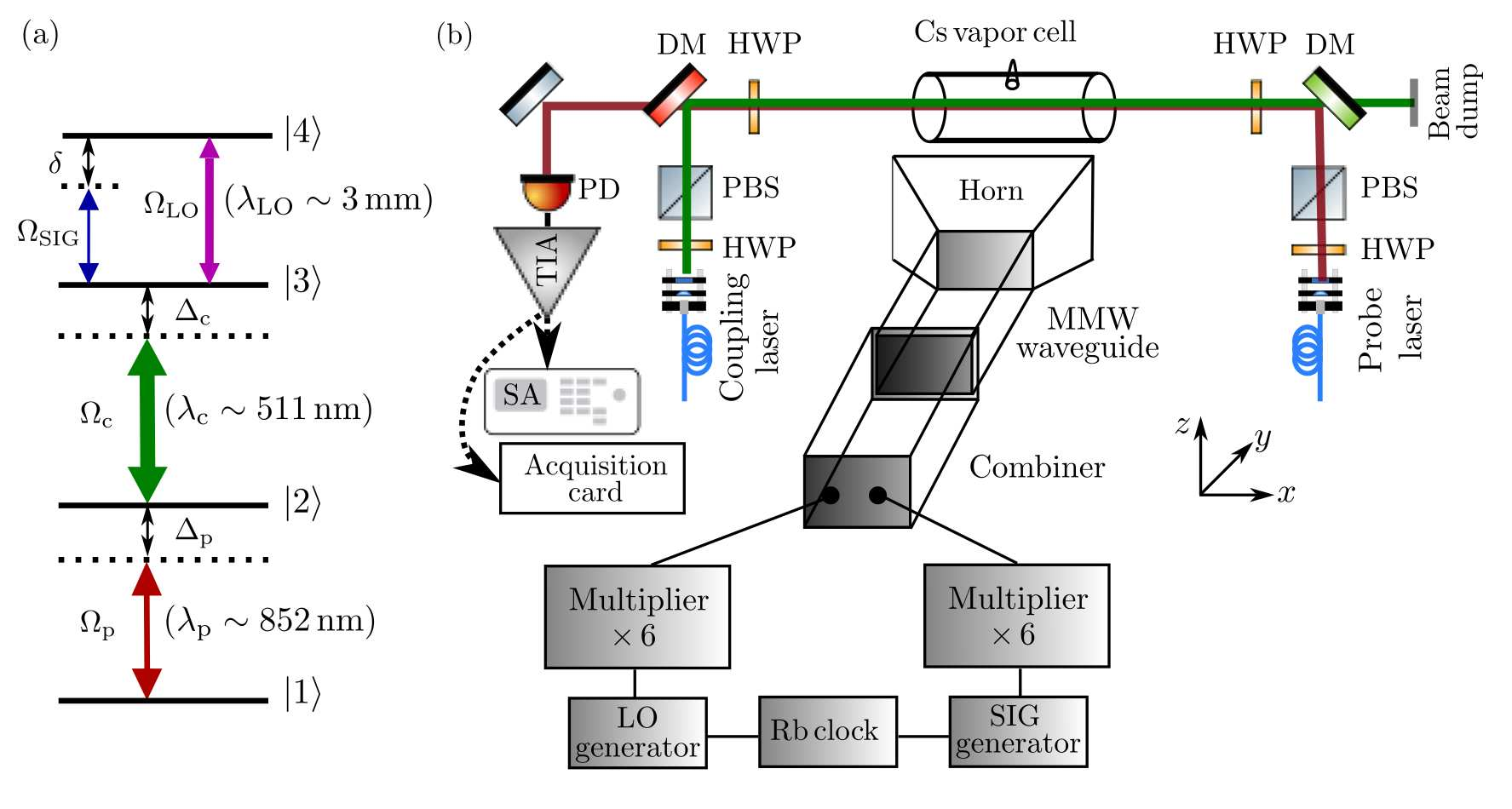}
\caption{Two-photon Rydberg excitation in $^{133} \mathrm{Cs}$ vapor cell : (a) four-level energy diagram with ground state $\ket{1} =  \ket{6^{2}S_{1/2}}$, first excited state $\ket{2} = \ket{6^{2}P_{3/2}}$, Rydberg states $\ket{3} = \ket{37^{2}S_{1/2}}$ and $\ket{4} = \ket{36^{2}P_{3/2}}$. The corresponding Rabi frequencies and detunings of the 852-nm probe laser and 511-nm coupling laser are, respectively, $\Omega_{\mathrm{p}}$~($\Omega_{\mathrm{c}}$) and $\Delta_{\mathrm{p}}$~($\Delta_{\mathrm{c}}$). A strong MMW~($\lambda_{\mathrm{LO}} \sim 3~\mathrm{mm}$) resonant local oscillator~field~($\Omega_{\mathrm{LO}}$) and a weak MMW signal~field~($\Omega_{\mathrm{SIG}}$) with RF detuning $\delta$ are coupling the two Rydberg states $\ket{3}$ and $\ket{4}$. (b)~Setup for superheterodyne reception. HWP: half-waveplate, QWP: quarter waveplate, DM:  dichroic mirror,  PD: photodiode, TIA:  transimpedance amplifier, PBS: polarizing beam splitter, SA: spectrum analyzer.
}
\label{fig:2}
\end{figure*}

\begin{figure*}[t]
\includegraphics[width=18cm]{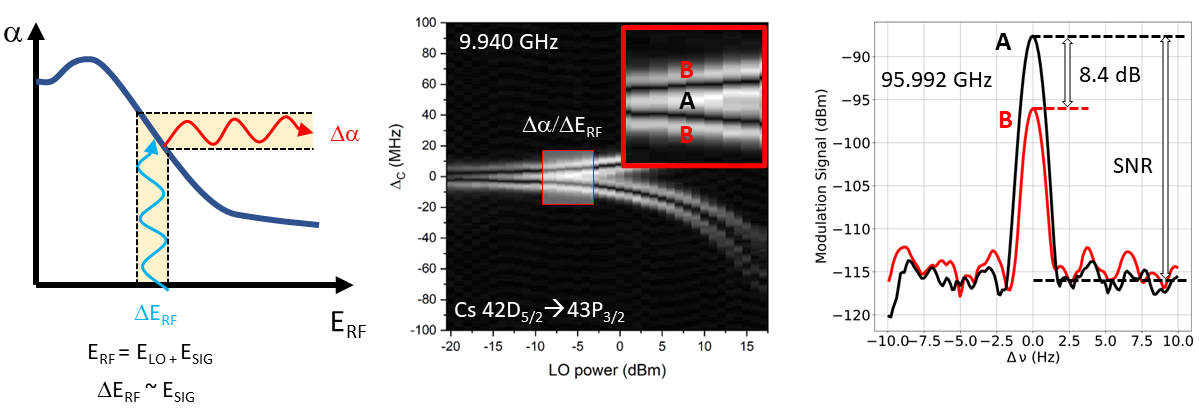}
\caption{Responsivity of atomic receivers to microwave (MW) and millimeter-wave (MMW) signals.  Left: illustration of a typical EIT probe absorption coefficient $\alpha (E_{RF}, \Delta_c)$ at a fixed EIT coupling detuning, $\Delta_c=0$, versus RF field, $E_{RF}$. $\Delta\alpha/\Delta E_{RF}$ at $\Delta_c$ is the relevant atomic figure of merit for responsivity. Middle: measurement of $
\Delta\alpha/\Delta E_{RF} (E_{RF}, \Delta_c)$ with the atomic receiver for a 9.940~GHz MW field. Right: Signal-analyzer (SA) spectra of
Rydberg-receiver output signals on operating points A and B for an equivalent measurement of a ~95.992512~GHz MMW signal on the Cs $37S_{1/2}\leftrightarrow 36P_{3/2}$ Rydberg transition with $E_{\mathrm{LO}} = 0.51$~V/m and $E_{\mathrm{SIG}} = 513~\mathrm{\mu V/m}$. The SA frequency, $\Delta\nu$, is shown relative to the IF beatnote frequency of 72~kHz. Operating point A has a 8.4~dB higher response than operating point B.}
\label{fig:3}
\end{figure*}

To address this problem, we implement an erbium-doped fiber optical frequency comb (OFC) to transfer the linewidth and stability of a high-finesse optical cavity to our continuous-wave (CW) Rydberg lasers. These are key elements to advanced metrology applications such as optical clocks~\cite{Udem.2002, Nicholson.2012, Ye.2023}, time and frequency transfer~\cite{Droste.2013,Ellis.2021} as well as high-resolution spectroscopy~\cite{Collombon.2019, Grinin.2020}. The Rydberg comb laser system and stabilization scheme are depicted in Fig.~\ref{fig:1}. The femtosecond frequency comb laser is a mode-locked fiber laser at a wavelength of 1550~nm. The repetition frequency ($f_{\mathrm{rep}}$) of this laser is 200~MHz and can be fine-adjusted over a range of a kHz. The $f_{\mathrm{rep}}$-value is stabilized  using a heterodyne beat-note detector for a comb line and a narrow-linewidth ($<$~2~Hz) CW reference laser at 1550~nm, which itself is frequency-stabilized, here, to an ultra-stable high-finesse reference cavity made of ultra-low-expansion (ULE) glass. The 1550-nm/comb optical beat note is amplified, filtered and phase-locked to a 10-MHz reference signal from a Rubidium clock via a direct digital synthesizer (DDS). The $f_{\mathrm{rep}}$ error signal is fed back to piezo-electric tranducers (PZT) that control the laser resonator length.
The locked $f_{\mathrm{rep}}$-value is recorded using a frequency counter. A separate optical beat-note detector allows for carrier-envelope offset frequency ($f_{\mathrm{CEO}}$) detection and stabilization via a standard $f- 2f$ interferometer~\cite{Cundiff.2000}. The optical beat note is amplified, filtered and phase-locked to the Rb clock by another DDS, with the error signal fed back to the laser pump current. The locked $f_{\mathrm{CEO}}$ also is recorded using a frequency counter. Following the OFC stabilization, both 852-nm and 1020-nm lasers are phase-locked to the OFC using two additional heterodyne beat note detectors between the OFC and the lasers. In this way, the $10^{-14}$-level relative frequency stability of the 1550-nm reference laser is transferred to the frequencies of the 852-nm and 510-nm Rydberg-EIT lasers used in the atomic receiver.

\section{Millimeter-wave atomic superheterodyne receiver}

The four-level cesium ($^{133}$Cs) atom energy-level structure relevant to this work is illustrated in Fig.~\ref{fig:2}(a). The four atomic energy levels  $\ket{1}$, $\ket{2}$, $\ket{3}$ and $\ket{4}$ correspond, respectively, to the ground state $6^{2}S_{1/2}$, excited state $6^{2}P_{3/2}$, and Rydberg states $37^{2}S_{1/2}$ and $36^{2}P_{3/2}$. Rydberg EIT is obtained using a weak probe laser beam at 852~nm comb-locked on the resonant $D_{2}$ cycling transition   $\ket{6^{2}S_{1/2}, F=4} \rightarrow \ket{6^{2}P_{3/2}, F'=5}$ ($\lambda_{p} = 852.356~\mathrm{nm}$) and a strong coupling beam resonant with the $\ket{6^{2}P_{3/2}, F=5} \rightarrow \ket{37^{2}S_{1/2}}$ ($\lambda_{c} = 510.905~\mathrm{nm}$) Rydberg transition. The 852-nm probe laser is an external cavity diode laser (ECDL), and the 511-nm coupling laser is a frequency doubled 1020~nm~ECDL laser. The probe beam has a power of 158~$\mu$W, a full-width at half maximum (FWHM) of 636~$\mu$m and a Rabi frequency $\Omega_{\mathrm{p}}~/~2 \pi$~=~16.9~MHz. The coupling beam has a power of 60~mW, a beam FWHM of 1.1~mm and a Rabi frequency $\Omega_{\mathrm{c}}~/~2 \pi$~=~0.9~MHz. The optical Rabi frequencies are given for the peak power at the center of the Gaussian beams, providing an upper limit. A schematic of the experimental apparatus is shown in Fig.~\ref{fig:2}(b). The probe and coupling lasers are counter-propagated, collimated, and overlapped within a cylindrical glass vapor cell of length $L$~=~30-mm containing a room-temperature vapor of Cs atoms. The power of the transmitted probe beam is recorded with a photodiode, the signal of which is amplified and recorded either by a Spectrum Analyzer (SA) for frequency analysis or digitized in real-time for time-domain signal analysis. \\

On-resonant Autler-Townes (AT) splitting is used to calibrate the MMW E-fields of both LO and signal fields inside the vapor cell that are generated from the MMW transmission line. The fields are calibrated by recording AT-split Rydberg EIT spectra, scanning the coupling-laser detuning $\Delta_{\mathrm{c}}$, while the probe laser is stabilized to the $\ket{6^{2}S_{1/2}, F=4} \rightarrow \ket{6^{2}P_{3/2}, F'=5}$~$D_{2}$-cycling transition~(probe detuning~$\Delta_{\mathrm{p}} = 0$). The coupling-laser detuning, $\Delta_c$, is scanned by tuning the DDS associated with the 1020~nm-laser (see Fig.~\ref{fig:1}).\\
While useful to calibrate the LO and signal (SIG) $E$-fields,
$E_{\mathrm{LO}}$ and $E_{\mathrm{SIG}}$,
the measurement of AT splittings is not a viable approach for the sensing of weak $E$-fields field sensing, where the signal's atom-field interaction Rabi frequency is well below the EIT linewidth, rendering field-induced spectral changes too small to accurately resolve (see e.g. Fig.~\ref{fig:3}b). For weak-field measurements, the MMW atomic receiver incorporates a compact cesium atomic vapor cell and optical readout of RF-sensitive Rydberg states using two-photon Rydberg EIT and a MMW heterodyne (HET) architecture.  Figure~\ref{fig:2} shows the atomic MMW receiver setup.  The atomic MMW HET uses an LO field  $E_{\mathrm{LO}}$
as a frequency reference, with a frequency near-resonance with the Rydberg transition~~\cite{Anderson.2019}.
The LO field is superimposed in the atomic vapor with a weak SIG field $E_{\mathrm{SIG}}$ of interest at a frequency detuning $\delta$ from the LO frequency. The atoms respond to the time-varying field superposition oscillating at the difference frequency between $E_{\mathrm{SIG}}$ and $E_{\mathrm{LO}}$, or intermediate frequency~(IF), which is detected optically.
The $E_{\mathrm{LO}}$ and $E_{\mathrm{SIG}}$ fields are generated by two independent signal generators (SG), generating $\tilde{f}_{\mathrm{LO}}~=~15.998740$~GHz and $\tilde{f}_{\mathrm{SIG}}~=~15.998752$~GHz, both synchronized to a common 10-MHz reference from a Rubidium clock. Each SG field is passed through a $6\times$ multiplier to generate MMW fields at frequencies $f_{\mathrm{LO}}~=~95.992440$~GHz and $f_{\mathrm{SIG}}~=~95.992512$~GHz that are subsequently combined into a horn antenna and directed over-the-air to the atomic vapor. The emitted fields are linearly polarized along the $z$-direction. The horn antenna is placed at a distance $d$~=~100~mm from the center of the vapor cell in the antenna far-field $d>\frac{2 \, a^{2}}{\lambda_{\mathrm{RF}}}~=~88.5$~mm, where $a=11.757$~mm is the diagonal of the horn antenna and $\lambda_{\mathrm{RF}}=3.125$~mm is the MMW-field wavelength. After calibrating $E_{\mathrm{LO}}$ and $E_{\mathrm{SIG}}$, the IF beatnote $f_{\mathrm{IF}} = \delta = 72$~kHz is measured as described next.

\section{Responsivity}

Understanding the responsivity of atomic receivers to signal fields is a prerequisite to a systematic characterizations of their performance and their operation in applications.  In Figure~\ref{fig:3}, we characterize the responsivity of the atomic receiver to weak MW (10~GHz) and MMW (100~GHz) fields resonant with atomic Rydberg transitions. The left plot in Fig.~\ref{fig:3} illustrates the behavior of the probe absorption in the atomic EIT medium $\alpha(E_{\mathrm{RF}}, \Delta_{\mathrm{c}})$ as a function of an applied RF electric field of the form $E_{\mathrm{RF}}=E_{\mathrm{LO}}+E_{\mathrm{SIG}}$ with a laser detuning $\Delta_{\mathrm{c}}=0$. The highest sensitivity to a small change in the field, $\Delta E_{\mathrm{RF}}$, is achieved when $E_{\mathrm{LO}}$ is set to a value of maximum slope ({\sl{i.e.}}, an inflection point) on the curve $\alpha(E_{\mathrm{RF}}, \Delta_{\mathrm{c}}=0$). It is thus seen that the exact LO field value is critical because the LO must shift the operating point away from $E_{\mathrm{RF}}=0$ to a locus where the response to small field changes, $\Delta E_{RF}$, is maximal and linear, and where the dynamic range in $\Delta E_{\mathrm{RF}}$ is large.

We establish a figure of merit for the responsivity as $\Delta\alpha/\Delta E_{\mathrm{RF}}$ on the $(\Delta_{\mathrm{c}}, E_{\mathrm{LO}})$-plane.
In Figure~\ref{fig:3}~(middle), $\Delta\alpha/\Delta E_{\mathrm{RF}}$ is measured for the atomic receiver on the Cs~$\mathrm{42^{2}D_{5/2}}\leftrightarrow \mathrm{43^{2}P_{3/2}}$ Rydberg transition for near-resonant LO and SIG MW frequencies around 9.940~GHz as a function of $\Delta_{\mathrm{c}}$ and $E_{\mathrm{LO}}$ for a fixed, small $E_{\mathrm{SIG}}$. This responsivity plot for resonant RF has the shape of a rotated Y; we therefore refer to it as ``Y-map''. For the set of atomic-receiver parameters used, the best operating point with the highest $\Delta\alpha/\Delta E_{\mathrm{RF}}$ is located where the diagonal legs of the ``Y'' intersect, highlighted in the red box in Figure~\ref{fig:3}~(middle). The highest responsivity occurs at the operating point labeled A, which resides at said intersection point of the Y-legs with the $\Delta_{\mathrm{c}}=0$-line. Operating point A corresponds with a well-defined $E_{\mathrm{LO}}$-value that yields best signal response.
Several secondary operating points, labeled B on the Y-map, occur at the same $E_{\mathrm{LO}}$-value as A, but at symmetric non-zero detunings $\Delta_{\mathrm{c}}$ of approximately $\sim10$~MHz. The B-points have a responsitity of about 8.4~dB less than that of point A.  This is shown in Figure~\ref{fig:3}~(right) for the Cs~$\mathrm{37^{2}S_{1/2}} \leftrightarrow \mathrm{36^{2}P_{3/2}}$ Rydberg transition at a MMW frequency of 95.992512~GHz.

\section{Sensitivity and dynamic range}

\begin{figure}[t]
\includegraphics[width=7cm]{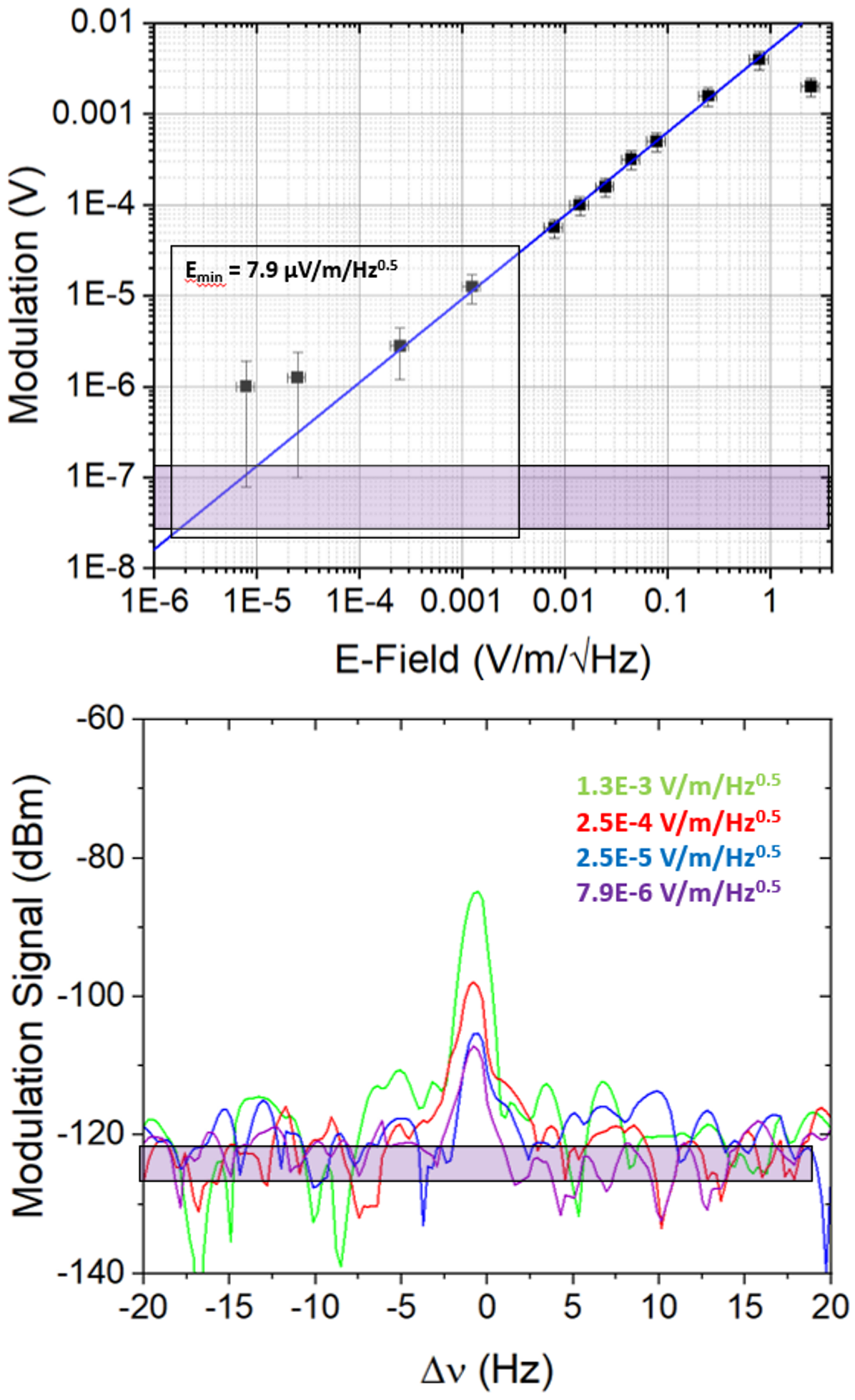}
\caption{Atomic receiver sensitivity and dynamic-range measurement for a 95.992512~GHz SIG field for a detuning $\Delta$f of the SIG field from the LO, equivalent to an in intermediate frequency (IF) of 72~kHz (IF = SIG - LO frequency). Top:
Receiver's output signal in RMS voltage
at the IF frequency versus applied MMW SIG electric field in units V/m/$\sqrt{{{\rm Hz}}}$.
Bottom: SA spectra in dBm centered at the 72-kHz IF and measured at the four lowest SIG fields of the top plot (values provided in the legend in units V/m/$\sqrt{{{\rm Hz}}}$).}
\label{fig:4}
\end{figure}

We employ our atomic sensor architecture and its characteristic responsivity to MMW fields to maximize the responsivity $\Delta\alpha/\Delta E_{\mathrm{RF}}$, under
the constraint that the signal-to-noise ratio (SNR) of its IF output, as defined in Fig.~\ref{fig:3}~(c), also is maximal. The sensitivity is maximized on operating point A by a parameter optimization on all optical powers, optical detunings $\Delta_{\mathrm{c}}$ and $\Delta_{\mathrm{p}}$, MMW LO power, and RF detuning $\delta$. This allows us to establish a minimum $E$-field sensitivity.
In addition, we establish the linear dynamic range of the MMW atomic receiver at operating point A for a
95.992512~GHz W-band SIG test field.

Figure~\ref{fig:4} shows the signal analyzer RMS voltage as a function of 95.992512~GHz SIG electric field for the Cs~$\mathrm{37^{2}S_{1/2}} \leftrightarrow \mathrm{36^{2}P_{3/2}}$ MMW transition of Cs (dipole moment $\mathcal{D} = 1150$~ea$_0 \, \times  \, 0.4714$, where 1150~ea$_0$ is the radial and 0.4714 the angular matrix element). We measure a minimum detectable MMW $E$-field given by the intercept of the blue line, which extends the linear response into the low-field regime to the noise floor. The field value of the intercept at the noise floor yields a minimum detectable MMW field, which establishes the sensitivity figure of the Rydberg MMW receiver. Here, we find this to be $E_{\mathrm{min}}=7.9~\mathrm{\mu V/m}$ for a 1-second integration time, limited by residual optical technical noise. Figure~\ref{fig:4} further shows that the
receiver exhibits a linear response over a 35-dB-wide range in MMW electric field, starting at
$\sim$~0.3~mV/m/$\sqrt{{\rm{Hz}}}$ and beginning to saturate at $\sim$~1~V/m/$\sqrt{{\rm{Hz}}}$, where the voltage levels off as a function of field. This finding corresponds to a linear dynamic range of about 70~dB in MMW power.

\section{Selectivity}

\begin{figure}[h!]
\centering
\includegraphics[width=8cm]{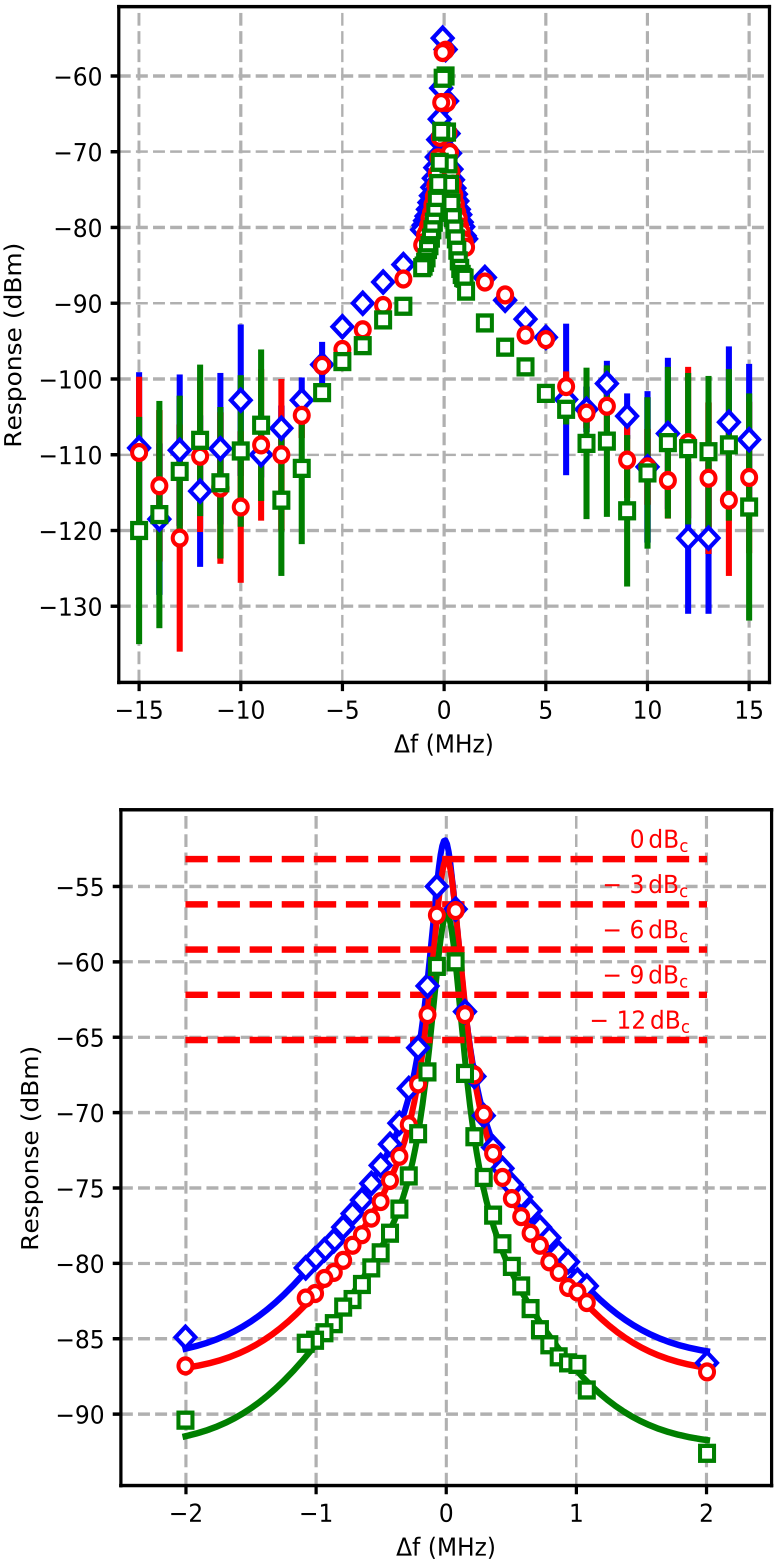}
\caption{Atomic receiver response to signal frequency (SIG frequency detuning $\Delta f$ from SIG=LO, or IF frequency). Top: Response over a $\Delta f$ range from $-$15~MHz to $+$15~MHz (IF = SIG - LO frequency), for LO-frequency 95.992440~GHz resonant with a Rydberg MMW transition. Bottom: Atomic receiver responsivity over an IF range of $-$2~MHz to $+$2~MHz.  Response curves are shown for coupling laser powers of 48~mW (blue diamonds), 38~mW (red circles), and 28~mW (green squares).  The response (pass band filter) curves are empirically fitted to Voigt functions (solid lines) before extraction of responsivity attenuation levels. The fitted Gaussian and Lorentzian widths in the Voigt fits are $\Gamma_{\mathrm{G}} / 2 \pi=\mathrm{\{547 , 522 , 537\}~kHz}$ and $\Gamma_{\mathrm{L}}/ 2 \pi=\mathrm{\{309 , 324 , 337\}~kHz}$, for respective coupling laser powers listed above. The red dashed lines show the extraction of the IF bandwidth cutoff frequencies for $\{3, 6, 9, 12\}$~dB attenuation levels at a coupling power of 38 mW.}
\label{fig:5}
\end{figure}

The selectivity of a traditional RF receiver is largely dictated by the signal response characteristics of the antenna and front-end mixer, amplifier, and filtering electronics.  For the atomic receivers with external LO fields, a novel type of selectivity filter is naturally provided by the
quantum-optical EIT response of the atoms to the detected fields within the receiver's atomic vapor cell, in which the SIG field first interacts and interferes with the LO field in the atomic medium.
Here, the selectivity is characterized from the readout of the atomic HET comprising the Rydberg-atom EIT and LO field system, which serves as the first frequency selector for the incoming SIG wave.

To establish a selectivity metric for an atomic receiver's ability to naturally reject out-of-band interference, we define a SIG rejection figure as the receiver's detected SIG power level in dB relative to the detected SIG power level at near-zero IF (where SIG and LO frequencies are approximately equal) at SIG frequency offsets $\vert \Delta f \vert /f=10^{-4}$, $10^{-5}$, and $10^{-6}$. Here, $\vert \Delta f \vert$ is the detuning of the SIG field from the LO, equivalent to an intermediate frequency (IF), and $f$ is the carrier (LO) frequency.  Table~\ref{Table:1} lists the measured SIG rejection values in dB at the defined $\vert \Delta f \vert /f$ frequency offsets.

As an additional metric to characterize the atomic receiver's selectivity we define 3~dB, 6~dB, 9~dB and 12~dB IF bandwidths, or $\vert \Delta f \vert$ cutoff frequencies, at which the receiver's detected SIG output power relative to the detected SIG output power at near-zero IF drops by 3~dB, 6~dB, 9~dB and 12~dB. In the following pilot test of these atomic-receiver metrics, we extract the values of the metrics from a measurement of the Rydberg receiver's output power as a function of IF frequency over a range of $\pm~$15~MHz, for an LO (carrier) with frequency $f_{\mathrm{LO}} = f = 95.992440$~GHz (which is resonant
with a strong Rydberg transition).

Figure~\ref{fig:5}~(top) shows the spectrum-analyzer power readout from the atomic receiver (in dBm) at a signal electric field $E_{\mathrm{SIG}}= 59.6~\mathrm{mV/m}$ as a function of IF $\Delta f$ over a range from $-$15~MHz to $+$15~MHz. The IF = SIG frequency - LO frequency. The LO frequency $f_{\mathrm{LO}}=95.992440$~GHz is resonant with the Cs~$\mathrm{37^{2}S_{1/2}} \leftrightarrow \mathrm{36^{2}P_{3/2}}$ Rydberg transition, and the LO field is $E_{\mathrm{LO}}= 0.65~\mathrm{V/m}$. The measurement is performed for three different coupling powers, $\mathrm{P_{c}}=\{48, 38, 28\}$~mW, and Rabi frequencies, $\Omega_{\mathrm{c}}/2 \pi=\{0.82, 0.73, 0.63\}$~MHz, respectively. Figure~\ref{fig:5}~(bottom) shows a higher-resolved selectivity measurement for the IF ranging from $-$2~MHz to $+$2~MHz. A Voigt profile was empirically found to provide the best fit to the measured curves. The Gaussian and Lorentzian linewidths from fits for coupling powers $\mathrm{P_{c}}=\{48, 38, 28\}$~mW are found to be $\Gamma_{\mathrm{G}} / 2 \pi=\mathrm{\{547 , 522 , 537\} \,\, kHz}$ and $\Gamma_{\mathrm{L}} / 2 \pi=\mathrm{\{309 , 324 , 337\} \,\, kHz}$, respectively. The fit functions are then employed to extract the metrics defined above.

\begin{table}[h!]
\begin{center}
\caption{Signal rejection in dB as a function of $\vert \Delta f \vert /f$.}
\label{Table:1}
\begin{tabular}
{| m{2.5cm} |  m{1.3cm} | m{1.3cm} |  m{1.3cm} | }
\hline
 $| \Delta f |/f$ & $10^{-6}$ & $10^{-5}$ & $10^{-4}$   \\
 \hline
Rejection (dB) & 8 & 29 & 53  \\
 \hline
\end{tabular}
\end{center}
\end{table}

\begin{table}[h!]
\begin{center}
\caption{IF bandwidths at selected coupling-laser powers.}
\label{Table:2}
\begin{tabular}
{| m{1.3cm} ||  m{1.3cm} || m{1.3cm} ||  m{1.3cm} ||  m{1.3cm} |}
\hline
 \textbf{Power (mW)} & \textbf{3-dB (kHz)} & \textbf{6-dB (kHz)} & \textbf{9-dB (kHz)} & \textbf{12-dB (kHz)} \\
 \hline
 28 & 129.8 & 198.6 & 266.8 & 344.7 \\
 \hline
 38 & 127.4 & 195.6 & 263.9 & 343.2 \\
 \hline
 48 & 123.3 & 189.8 & 257.2 & 336.6 \\ [1ex]
 \hline
\end{tabular}
\end{center}
\end{table}

\begin{table}[h!]
\caption{Shape factors at selected coupling-laser powers.}
\label{Table:3}
\begin{tblr}{| c ||  c || c|| c |}
\hline
 \textbf{Power (mW)} & \textbf{$\mathrm{\frac{BW_{10dB}}{BW_{3dB}}}$} & \textbf{$\mathrm{\frac{BW_{20dB}}{BW_{3dB}}}$} & \textbf{$\mathrm{\frac{BW_{30dB}}{BW_{3dB}}}$} \\
 \hline
 28 &  2.3 & 5.5 & 16.3 \\
 \hline
 38 & 2.3 & 5.8 & 17.7  \\
 \hline
 48 & 2.3 & 6.2 & 19.1  \\
 \hline
\end{tblr}
\end{table}

The receiver signal selectivity, defined as the ratio comparing the detected wanted signal frequency (SIG=LO) and an unwanted signal frequency (SIG-LO$=\Delta f\ne$ 0) in dB for selected frequency off-sets $\vert \Delta f \vert /f=10^{-6}, 10^{-5},$ and $10^{-4}$ are shown in Table~\ref{Table:1}. The effect of coupler-power change in Fig.~\ref{fig:5} largely amounts to a global vertical shift by fixed and small dB-values, and the results in Table~\ref{Table:1} do not depend on coupler-laser power. The signal rejection improves from about 8~dB at $\vert \Delta f \vert /f = 10^{-6}$ (corresponding to a full IF width of 200~kHz) to as much as 53~dB $\vert \Delta f \vert /f = 10^{-4}$ (corresponding to a full IF width of 20~MHz). These results highlight the potential for high frequency selectivity and filtering naturally afforded by atomic receivers, prior to any electronic analog or digital signal processing.

IF bandwidths at $\{3, 6, 9, 12\}$~dB attenuation levels are tabulated in Table~\ref{Table:2}.  Over the coupler-power range investigated, little change is observed in the lineshape of the IF response curve in Fig.~\ref{fig:5}. As such, the data in Table~\ref{Table:2} do not exhibit a significant dependence on coupler power. The results in Table~\ref{Table:2} show that the presently configured millimeter-wave atomic receiver provides high selectivity and narrow bandwidth.

Another figure of merit related to selectivity is the shape factor of the atom radio receiver, defined as the ratio of the bandwidth (BW) of the filter response curve relative to the 3-dB bandwidth, i.e. shape factor $\mathrm{BW_{30dB}/BW_{3dB}}$ is the IF bandwidth of the response curve at 30~dB signal rejection divided by the BW at 3-dB (or 3-dB BW).  Table~\ref{Table:3} shoes measured shape factors for 10-dB, 20~dB, and 30~dB for the different coupling-laser powers. In the present case the shape factor is about 19:1 for $\mathrm{BW_{30dB}/BW_{3dB}}$ and reduces at lower dB and coupler power.  The shape factor of the atom radio filter may be reduced or changed by tuning or varying other parameters to make appropriate use of the overall available RF signal spectrum in the presence of unwanted signals that must be blocked out.


\begin{figure}[h!]
\centering
\includegraphics[width=8cm]{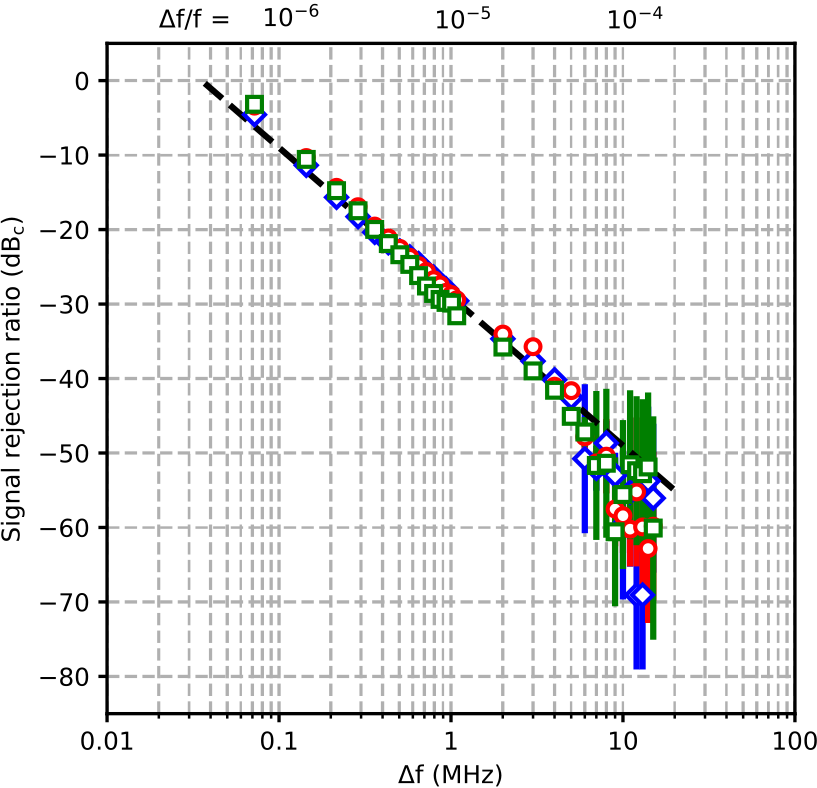}
\caption{Atom receiver filter roll-off showing signal rejection ratio in $\mathrm{dB_{c}}$ versus signal frequency detuning $\Delta f$. Notably, at large $|\Delta f|$ the atomic receiver filter roll-off exhibits a fixed slope of -2, as indicated by the black dashed line.}
\label{fig:6}
\end{figure}

Finally, in Fig.~\ref{fig:6} we show the receiver's filter roll-off in $\mathrm{dB_{c}}$ as a function of $\Delta f$, on a log scale. The data points are equivalent with those in Fig.~\ref{fig:5}~(top), minus the response at zero IF, $\Delta f = 0$, taken from the Voigt profile fit. Figure~\ref{fig:6} emphasizes the signal rejection scaling behavior at IF larger than about 100~kHz. The data in Fig.~\ref{fig:6} follow a linear trend with a slope of about -2, showing that the rejection scales as $|\Delta f|^{-2}$. This accords with our finding that Voigt profiles deliver good empirical fits to the selectivity spectra. At large argument, the decay of a Voigt profile comes from the wide Lorentzian wings of the profile, which drop off as $|\Delta f|^{-2}$ at large  $|\Delta f|$.

\section{Conclusion}

We have described and demonstrated an atomic receiver for millimeter-wave signal detection.  We have presented a comb-stabilized Rydberg laser system and atomic heterodyne receiver architecture.  We investigated the receiver responsivity to microwave and millimeter-wave signal fields as a function of receiver reference LO field and coupler laser detuning parameters. An optimal operating point was identified within this receiver parameter space; secondary operating points that are down in responsivity by 8~dB were also identified. We performed a detailed characterization of the atomic millimeter-wave receiver, achieving 7.9~$\mu$V/m/$\sqrt{\mathrm{Hz}}$ minimum field sensitivities and a $>$70~dB linear dynamic range, limited by technical noise and non-linear atomic response in the strong atom-field interaction regime, respectively.  We established a metric for selectivity in atomic receivers as well as a measurement protocol, which we employed to characterize the selectivity for the atomic millimeter-wave receiver.  Selectivity, rejection figures, IF bandwidths, cutoff frequencies, filter roll-off and shape factors have been quantified.  This work represents an important advance in future studies and applications of atomic receiver science and technology, establishes a new state of the art in millimeter-wave sensitivity and performance metrics with Rydberg atomic receivers, and in weak MMW and high-frequency signal detection.

\section{Acknowledgements}
The authors thank Rachel Sapiro for early contributions to this work.  This work was supported by Rydberg Technologies Inc. and, in part, by the Defense Advanced Research Projects Agency (DARPA) SAVaNT Program under agreement No. HR00112190065. The views and conclusions contained in this document are those of the authors and should not be interpreted as representing the official policies, either expressed or implied, of the U.S. Government.



%

\end{document}